\documentstyle{EuroPhys}
\input EuroMacr
\input epsf
\begin{document}
\newcommand{\dd}{{\rm d}} \newcommand{\td}{{\bf\widehat{m}}}
\euro{}{}{}{} \Date{} \shorttitle{}
\title{Coupling between membrane tilt-difference and dilation:\\
a new ``ripple'' instability and multiple crystalline\\
  inclusions phases} \author{J.-B. Fournier\,\footnote{e-mail:
    jbf@turner.pct.espci.fr}} \institute{
  Laboratoire de Physico-Chimie Th\'eorique,\\
  \'Ecole Sup\'erieure de Physique et de Chimie Industrielles de la
  Ville de Paris,\\
  10 rue Vauquelin, F-75231 Paris C\'edex 05, France\\
  } \rec{}{} \pacs{ \Pacs{87}{22Bt}{Membrane and subcellular physics
    and structure} \Pacs{34}{20$-$b}{Interatomic and intermolecular
    potentials and forces, potential energy surfaces for collisions} }
\maketitle
\begin{abstract}
  A continuum Landau theory for the micro-elasticity of membranes is
  discussed, which incorporates a coupling between the bilayer
  thickness variation and the difference in the two monolayers' tilts.
  This coupling stabilizes a new phase with a rippled micro-structure.
  Interactions among membrane inclusions combine a dilation-induced
  attraction and a tilt-difference--induced repulsion that yield $2D$
  crystal phases, with possible coexistence of different
  lattice spacings for large couplings.  Inclusions favoring
  crystals are those with either a long-convex or a short-concave
  hydrophobic core.
\end{abstract}

Lipid molecules in water spontaneously form fluid bilayers in which
hydrocarbon tails are shielded from contact with
water~\cite{Isr},~\cite{Saf}. In nature, lipid membranes constitute
the walls surrounding living cells, and generally host a large number
of (protein) inclusions~\cite{Bio}. Self-assembled surfactant
membranes can form various phases, {\em e.g.}, lamellar ($L_\alpha)$,
vesicular ($L_4$), or sponge ($L_3$) phases. At low temperatures, the
molecules tilt relative to the membrane normal, forming the
$L_{\beta'}$ phase which has an internal degree of freedom similar to
that of the liquid crystalline smectic-$C$ phase. The coupling between
tilt and membrane curvature can produce a shape instability yielding
the $P_{\beta'}$ or ``ripple''
phase~\cite{Tardieu73},~\cite{Lubensky93}. Recently, a new degree of
freedom has been introduced by Seifert {\em et al\,}.: a {\em
  tilt-difference} between the two membrane monolayers.  The coupling
between tilt-difference and membrane curvature can produce
instabilities yielding rippled phases, bilayer tubules and
bicontinuous phases~\cite{Seifert96}, very much like in the case of
membranes containing nematogens or anisotropic
inclusions~\cite{ani},~\cite{ani2}.

In this letter, we study the effects of the coupling between
tilt-difference and membrane dilation in ordinary $L_\alpha$
membranes. We find a new type of ``ripple'' phase that has not yet
been evidenced. The tilt-difference and dilation modes are generically
excited by protein inclusions having a {\em convex}\, or a {\em
  concave}\, shape, or a hydrophobic core with a thickness different
from that of the bilayer.  We calculate and discuss the short-range
interactions between such hosts, which form the most general up-down
symmetric inclusions.

\begin{figure}
\centerline{\hspace{2cm}\epsfxsize=10cm\epsfbox{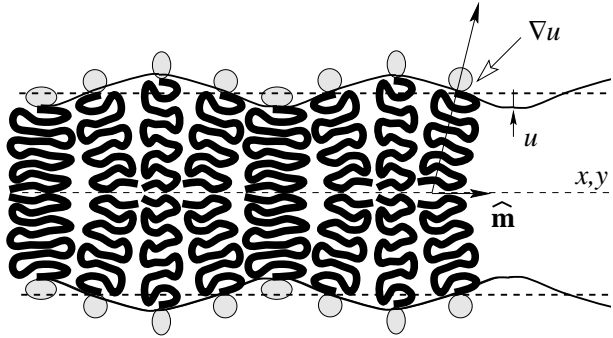}}
\caption{Lipid membrane subject to the ripple instability produced by
  the coupling $c\,\td\cdot\nabla u$ between the dilation\,($u$) and
  tilt-difference\,($\td$) modes.\vspace{-5pt}}
\label{inst}
\end{figure}

At lengths scales large compared with molecular dimensions, membranes
are traditionally described by the shape $h({\bf r})$ of their
midsurface, with ${\bf r}=(x,y)$.  Their elasticity is governed by the
Helfrich curvature Hamiltonian
$\frac{1}{2}(\kappa+\bar\kappa)(\nabla^2h)^2 -\frac{1}{2}\bar\kappa\,
(\partial_i\partial_jh)(\partial_i\partial_jh)$~\cite{Helfrich73}.
There is no $(\nabla h)^2$ surface tension term: membranes are
self-assembled systems that optimize their area per molecule (or their
coarse-grained area at larger scales.)  We use here a different
approach since we are interested in the {\em microscopic} elasticity
of the bilayer~\cite{Marcelja76}-\cite{Aranda96}. To simplify, we only
consider membranes perfectly symmetric with respect to their
midsurface (a general theory treating the two monolayers independently
will be presented elsewhere~\cite{moi}.) We develop a phenomenological
Landau theory for the membrane dilation $u({\bf r})$ (thickness
variation), and the molecular tilt-difference $\td({\bf r})$. The
latter is defined as half the sum of the projections onto the $(x,y)$
plane of the unit vectors, oriented tail-to-head, parallel to the
molecules in the two monolayers (fig.~\ref{inst}). The model free
energy involves all quadratic terms and first order derivatives that
satisfy rotational symmetry:
\begin{equation}
  \label{nrj}
  f\!=\!
  \frac{1}{2}B\,u^2+
  \frac{1}{2}\lambda\left(\nabla u\right)^2+
  c\,\td\cdot\nabla u+
  \frac{1}{2}t'\,\td^2+
  \frac{1}{2}K'_1\,(\nabla\cdot\td)^2+
  \frac{1}{2}K'_2\,(\nabla\times\td)^2\,.
\end{equation}
Due to the tendency of the molecules to orient perpendicular to the
chain-water interface, we expect $c>0$.  For biological membranes, the
typical energy and length scales are given by $\kappa\simeq25\,k_{\rm
  B}T$ ($10^{-12}$\,erg) and $\xi_0\simeq20$\,\AA (monolayer
thickness)~\cite{Isr}.  This yields
$B\approx\kappa/\xi_0^4\simeq6\times10^{14}$~erg~cm$^{-4}$ which
agrees with the experimental value of the area-stretching coefficient
$B(2\xi_0)^2\simeq100$~erg~cm$^{-2}$~\cite{Isr}.  The dilation term
$\frac{1}{2}Bu^2$ has contributions from the water--chain\,(oil)
surface tension and from the polymer-like stretching of the chains.
The dilation-gradient term $\frac{1}{2}\lambda(\nabla u)^2$ therefore
originates from the extra cost of {\em modulating} the stretching of
the chains and increasing the water--chain area. Hence, we expect
$\lambda\approx B\xi_0^2\simeq25$~erg~cm$^{-2}$, dimensionally,
whether or not the (macroscopic, effective) tension of the membrane
vanishes. On this point, our model disagrees with those of Dan {\em et
  al.}~\cite{NoteDan}. On the same basis, the other coefficients $c$,
$t'$ are expected to be $\approx\kappa/\xi_0^2$ and the $K'_i$'s are
expected to be $\approx\kappa$. Neglecting higher-order gradient terms
in~(\ref{nrj}) is in fact only justified when $B$ is small, which is
the case in the vicinity of a transition to a more ordered $L_\beta$
or $L_{\beta'}$ phase. For a microscopic $\sqrt{\lambda/B}$, we expect
nonetheless to get qualitatively correct results.

\vspace{\baselineskip}
{\em ``Ripple'' phase}.\ --\ The linear stability of the flat membrane
can be studied in the reciprocal space, where the free-energy density
per mode associated with~(\ref{nrj}) can be diagonalized to
\begin{equation}
\frac{1}{2}B(k)\left| u_{\bf k}-{\rm
i}\frac{c\,k}{B(k)}\widehat{m}^{\parallel}_{\bf k}\right|^2+
\frac{1}{2}\frac{N(k)}{B(k)}\left|\widehat{m}^{\parallel}_{\bf k}\right|^2+
\frac{1}{2}t'(k)\left|\widehat{m}^{\perp}_{\bf k}\right|^2\,,
\end{equation}
where $\widehat{m}^{\parallel}_{\bf k}$ and $\widehat{m}^{\perp}_{\bf
  k}$ are the projections of the tilt-difference Fourier component
$\td_{\bf k}$ parallel and perpendicular to ${\bf k}$, respectively,
and $B(k)=B+\lambda k^2>0$, $t'(k)=t'+K'_2k^2>0$,
$N(k)=(t'+K'_1k^2)(B+\lambda k^2)-c^2k^2$. The stability of the flat
membrane is dictated by the sign of $N(k)$. When the latter is
positive, the minimum energy corresponds to $u_{\bf k}=\td_{\bf k}=0$
and the flat membrane is stable. Defining the dilation-, the
tilt-difference-- and the coupling-characteristic lengths, as
\begin{equation}
\xi=\sqrt{\frac{\lambda}{B}}\,,\quad
\xi'=\sqrt{\frac{K'_1}{t'}}\,\quad{\rm and}\quad
\ell=\frac{c}{2\sqrt{Bt'}}\,,
\end{equation}
respectively, $N(k)<0$ is equivalent to
$\xi^2\xi'^2k^4+(\xi^2+\xi'^2-4\ell^2)k^2+1<0$. Therefore the
instability occurs for $\ell>(\xi+\xi')/2$ at a nonzero wavevector
$k_c=\left(\xi\xi'\right)^{-1/2}$, the dilation and tilt-difference
modulations being in quadrature (fig.~\ref{inst}).

\vspace{\baselineskip}

{\em Short-range interactions among inclusions}.\ --\ Understanding
the membrane-mediated interactions among inclusions has recently
attracted much interest. Conical inclusions coupling to the local
membrane curvature are subject to {\em long-range} elastic and Casimir
(fluctuation) forces~\cite{Goulian93}-~\cite{Dommersnes98} that add up
to the standard direct forces. On the other hand, {\em short-range}
interactions arise from local structural perturbations: integral
proteins have a central hydrophobic region spanning the hydrophobic
core of the membrane and two polar extremities protruding outside.
Any thickness mismatch between the protein and the bilayer
hydrophobic regions results in a thickness perturbation that yields
membrane-mediated interactions~\cite{Owicki79}-\cite{Aranda96}.

\begin{figure}
  \centerline{\hspace{0cm}\epsfxsize=13.5cm\epsfbox{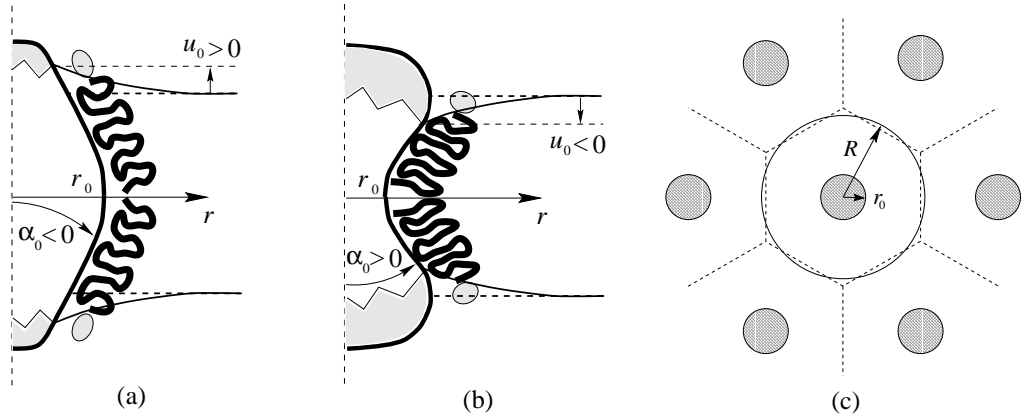}}
  \caption{Protein inclusions with thickness mismatch $u_0$ and
    tilt-difference angle $\alpha_0$. Inclusions forming $2D$ crystals
    have either a long-convex~(a) or a short-concave~(b) hydrophobic
    core. (c)~Wigner-Seitz cell approximated by a circle in an array
    of inclusions.}
  \label{incl_wigner}
\end{figure}

Here, we focus on the effects of the tilt-difference distortions that
are naturally excited by inclusions with a convex or concave
hydrophobic region. To simplify, we assume $\xi'\equiv\xi$ and hence
$\ell<\xi$ for the membrane's stability, and we restrict attention to
inclusions having revolution and up-down symmetry.  We denote by $u_0$
their hydrophobic thickness mismatch and by $\alpha_0$ their
tilt-difference angle (fig.~\ref{incl_wigner}a,b). Their mean field
interaction can be derived from the energy of the membrane equilibrium
structure, which obeys the Euler-Lagrange equations associated
with~(\ref{nrj}),
\begin{figure}
  \centerline{\hspace{0cm}\epsfxsize=14cm\epsfbox{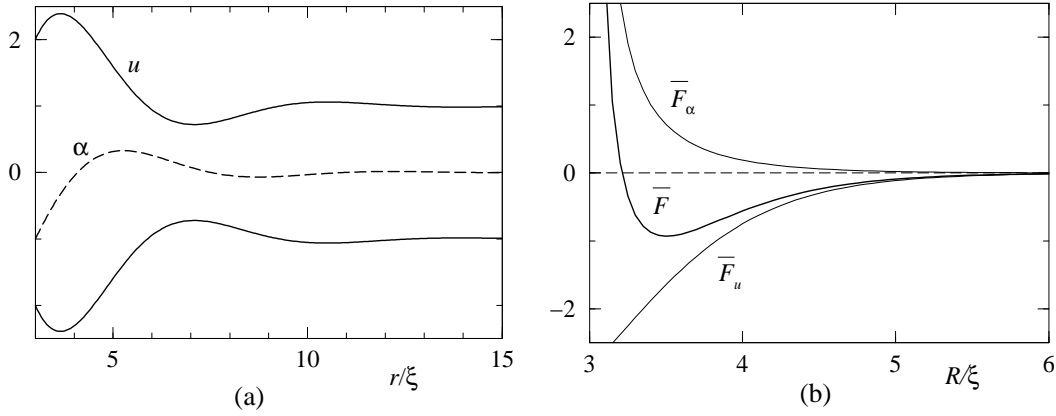}}
  \caption{(a)~Sketch of the membrane structure around an isolated
    inclusion with boundary dilation ($u$) and tilt-difference
    ($\alpha$) normalized to~$1$ ($r_0/\xi=3$, $x=-2$,
    $\phi=0.75\times\pi/2$).  (b)~Interaction energy $\overline F$ in
    units of $10\,k_{\rm B}T$ {\em vs.} separation for zero
    dilation--tilt-difference coupling ($r_0/\xi=3$, $x=0.5$,
    $\phi=0$).  \vspace{-6pt}}
  \label{membr_nrj}
\end{figure}
\begin{eqnarray}
  \label{uaa}
  &B\,u-\lambda\,\nabla^2u=
  c\,\nabla\cdot\td\,,&\\
  \label{uab}
  &t'\,\td-K'_1\,\nabla\left(\nabla\cdot\td\right)+
  K'_2\,\nabla\times\left(\nabla\times\td\right)=
  -c\,\nabla u\,.&
\end{eqnarray}
To study the physics of a collection of inclusions we consider the
Wigner-Seitz cell surrounding one inclusion with radius $r_0$ in a
lattice (fig.~\ref{incl_wigner}c) and we approximate it by a circle of
radius $R$~\cite{Owicki79},~\cite{Dan93}. The problem therefore
acquires revolution symmetry. In an hexagonal lattice, this amounts to
neglecting high-order Fourier harmonics ($6$, $12$, etc.), and in a
gas to considering that the first neighbors effectively screen the
other inclusions.  Assuming radial symmetry, $u=u(r)$ and
$\td=\alpha(r)\,{\bf\hat r}$, the equilibrium solutions are given by
the real part of
\begin{eqnarray}
  \label{1xiu}
  u(r)&=&\left[
    {\cal A}_1\,\,{\rm K}_0\!\left({\rm e}^{{\rm i}\phi}\,{r\over\xi}\right)+
    {\cal A}_2\,\,{\rm I}_0\!\left({\rm e}^{{\rm i}\phi}\,{r\over\xi}\right)
  \right]\times\sqrt{t'\over B}\,,\\
  \label{1xia}
  \alpha(r)&=&\left[
    {\cal A}_1\,\,{\rm K}_1\!\left({\rm e}^{{\rm i}\phi}\,{r\over\xi}\right)-
    {\cal A}_2\,\,{\rm I}_1\!\left({\rm e}^{{\rm i}\phi}\,{r\over\xi}\right)
  \right]\times{\rm i}\,,
\end{eqnarray}
where ${\rm i}=\sqrt{-1}$, the I's and the K's are modified Bessel function,
$\sin\phi=\ell/\xi$ ($\ell<\xi$), and ${\cal A}_1$, ${\cal A}_2$ are
two dimensionless complex constants that we determine form the
boundary conditions:
\begin{equation}
  u|_{r_0}=u_0\,,\quad
  \alpha|_{r_0}=\alpha_0\,,\quad
  \left.u'(r)\right|_R=0\,,\quad
  \left.\alpha\right|_R=0\,.
\end{equation}
The distortion energy $F$ stored in the region
$r_0\!<\!r\!<\!R$~\cite{note} can be transformed by integration by
parts into $F=\pi\left[\lambda\,r\,uu'
  +c\,r\,u\alpha+K'_1(r\,\alpha'+\alpha)\alpha\right]_{r_0}^R$, which
can be scaled to
\begin{equation}
  \frac{F}{\pi B\,r_0\,\xi\,u_0^2}=\overline{F}\left(
    x,\phi,\frac{r_0}{\xi},\frac{R}{\xi}
  \right)\,,
\quad{\rm with}\quad x=\frac{\alpha_0}{u_0\sqrt{B/t'}}\,.
  \label{eq:nor2}
\end{equation}

\begin{figure}
  \centerline{\hspace{0cm}\epsfxsize=13.5cm\epsfbox{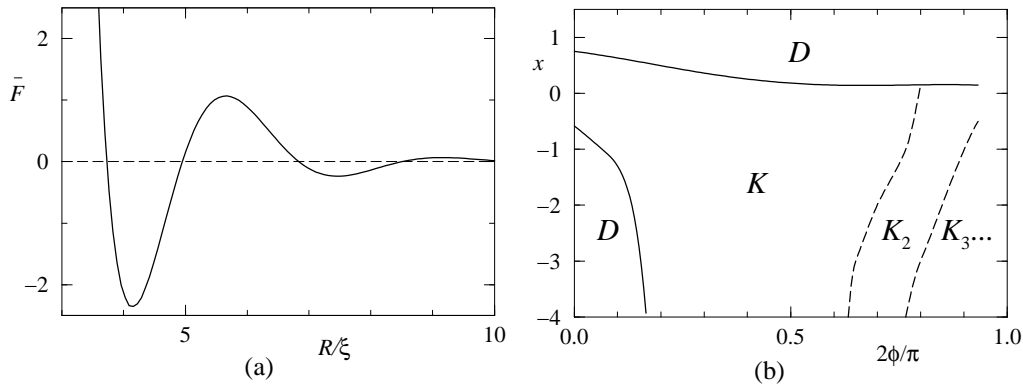}}
  \caption{(a)~Interaction energy $\overline F$ in
    units of $10\,k_{\rm B}T$ {\em vs.} separation for nonzero
    dilation-tilt-difference coupling ($r_0/\xi=3$, $x=-3$,
    $\phi=0.75\times\pi/2$). (b)~Phase diagram for a membrane with
    $\xi\simeq20\,\AA$ hosting inclusions with radius
    $r_0\simeq3\,\xi$.  ($D$) disordered. ($K$) crystal. ($K_n$)
    region where $n$ distinct stable or metastable crystalline phases
    are possible. For $x\simeq0$, the crystal spacing is so small that
    the inclusions effectively aggregate.\vspace{-12pt}}
  \label{2min_diag}
\end{figure}

For typical integral proteins, we expect $r_0\simeq3\,\xi$ with
$\xi\simeq20\,\AA$.  Both the dilation and the tilt-difference
distortion around an inclusions relax exponentially to zero on a
distance $\simeq100\,\AA$ ($\simeq5\,\xi$), as shown in
fig.~\ref{membr_nrj}a.  The coupling between dilation and
tilt-difference generates oscillations and overshoots.  For zero
coupling, $F$ can be splitted into a dilation-induced {\em attraction}
$F_u$ and a tilt-difference--induced {\em repulsion} $F_\alpha$ that
combine to yield a minimum for small enough
$x$~(fig.~\ref{membr_nrj}b). With the values of the material constants
previously discussed and typically
$u_0\simeq0.2\,\xi\,(\simeq4$~\AA), we expect $\alpha_0\simeq
x\times10$\,deg, and $\pi B\,r_0\,\xi\,u_0^2\simeq10\,k_{\rm B}T$,
which sets the magnitude of the interaction energy, and in particular
the depth of the well in fig.~\ref{membr_nrj}b. For large couplings,
interferences between the membrane oscillations yield several energy
minima (see the two wells of depths $\simeq25\,k_{\rm B}T$ and
$3\,k_{\rm B}T$ in fig.~\ref{2min_diag}a). 

Figure~\ref{2min_diag}b shows a typical phase diagram for a collection
of up-down symmetric inclusions. The criterion for a crystal phase is
the existence of a well deeper than $k_{\rm B}T$. Solid lines
indicate first-order phase transitions; dashed lines separate regions
where different metastable crystals with different lattice spacings can
exist. Remarkably, crystal phases occur preferentially for $x<0$, {\em
  i.e.}, for inclusions such as depicted in fig.~\ref{incl_wigner}a,b.

\vspace{\baselineskip}

{\em Discussion}.\ --\ The instability produced by the coupling
between tilt-difference and dilation defines a new ``ripple'' phase of
membranes.  The corresponding corrugation actually forms a {\em
  micro-structure}, since the undulation period compares with the
bilayer thickness (except maybe in the vicinity of a $L_\beta$ or
$L_{\beta'}$ phase). The amplitude of the corrugation, governed by the
bilayer dilation elasticity, will probably not exceed a few angstroms:
these ripples will be difficult to detect by electron microscopy or
STM techniques.  The present phase cannot be mistaken with the
$P_{\beta'}$ phase, which exhibits much larger (height) amplitudes
$\simeq45$~\AA~\cite{Zasadzinski88}.

Up-down symmetric membrane inclusions generally have a slightly convex
or concave hydrophobic core of thickness different from that of the
bilayer.  Due to the strong effective attraction between hydrophobic
parts, such inclusions will excite the coupled tilt-difference and
dilatation modes, which in turn will mediate short-range interactions
between them.  The thickness mismatch creates an energetic dilation
corona around the inclusions and yields an {\em attraction} between
like inclusions: no extra distortion occurs when the coronas overlap
since the boundary dilations match. The tilt-difference, however,
yields a {\em repulsion} between like inclusions: going from
$\alpha_0$ to $-\alpha_0$, it develops a strong
gradient when the coronas overlap.  Inclusion producing no
tilt-difference aggregate, while inclusions producing a nonzero
tilt-difference either repel one another or form $2D$ crystals. The
latter situation arises for small tilt-differences, or when the
dilation corona extends further than the tilt-difference corona
($\xi>\xi'$). In both cases, the attraction which dominates at
``large'' distances is overcome by the repulsion at short distances.

When the coupling between dilation and tilt-difference is large, the
distortions in the coronas exhibit damped oscillations and the
interparticle potential develops several minima. This implies the
possible coexistence of different crystals of inclusions having
different lattice spacings, and also the possibility of low density
crystals (separation between inclusions' boundaries
$\simeq180$~\,\AA~for the secondary minimum in fig.~\ref{2min_diag}).
The inclusions most likely to form $2D$ crystals are those with either
a {\em long-convex} or a {\em short-concave} hydrophobic core, {\em
  i.e.\/}, those disfavored from the point of view of the
$\td\cdot\nabla u$ coupling. This is because the gradient of $u$ being
more costly, the dilation corona extends (favoring ``long-range''
attraction), while due to the effective shift in the tilt-difference's
quadratic potential, the dilation corona shrinks (making the repulsion
occur only for smaller separations).  Conversely, short-convex and
long-concave inclusions have a dominant repulsion and should form
disordered phases.

In replacing the Wigner-Seitz cell by a circle, we have neglected
high-order Fourier harmonics. Taking them into account would probably
change the detail of the interaction potential, but not its essential
features. Real inclusions are not up-down symmetric: their average
conical shape effectively adds a monotonic
repulsion~\cite{moi}, which can be neglected provided the corresponding
angle is small compared to the tilt-difference angle.

%%%%%%%%%%%%%%%%%%%%%%%%%%%%%%%%%%%%%%%%%%%%%%%%%%%%%%%%%%%%%%%%%%%%%%


\begin{thebibliography}{99}
\bibitem{NoteDan} In~\cite{Dan93}-\cite{Aranda96}, the
  $(\nabla u)^2$ term would be restored if the lowest order term of
  the $u$-expansion would be considered a function
  $f_0(\Sigma,(\nabla\Sigma)^2)$ of the area per molecule {\em and its
    gradient}.
\end{thebibliography}
 \end{document}